\documentclass[twocolumn]{aastex63}

\newcommand{\oii}{[{\sc O\,ii}]}
\newcommand{\oiii}{[{\sc O\,iii}]}
\newcommand{\nii}{[{\sc N\,ii}]}
\newcommand{\sii}{[{\sc S\,ii}]}

\newcommand{\hii}{{\sc H\,ii}}

\newcommand\carter[1]{{\color{teal} #1}}

\newcommand\kms{$\mathrm{km\,s}^{-1}\,$}
\received{}
\revised{}
\accepted{\today}
\submitjournal{ApJ}

\usepackage[caption=false]{subfig}

\shorttitle{Machine Learning Approach to Spectroscopic Observations III}
\usepackage{amsmath}

\graphicspath{{./}{figures/}}
\usepackage{comment}
\begin{document}

\title{A Machine Learning Approach to Integral Field Unit Spectroscopy Observations: III. Disentangling Multiple Components in \hii{} regions}

\correspondingauthor{Carter L. Rhea}
\email{carter.rhea@umontreal.ca}

\author[0000-0003-2001-1076]{Carter L. Rhea}
\affiliation{D\'epartement de Physique, Universit\'e de Montr\'eal, Succ. Centre-Ville, Montr\'eal, Qu\'ebec, H3C 3J7, Canada}
\affiliation{Centre de Recherche en Astrophysique du Québec (CRAQ), Québec, QC, G1V 0A6, Canada}

\author[0000-0002-5136-6673]{Laurie Rousseau-Nepton}
\affiliation{Canada-France-Hawaii Telescope, Kamuela, HI, United
States}

\author[0000-0002-1755-4582]{Simon Prunet}
\affiliation{Canada-France-Hawaii Telescope, Kamuela, HI, United
States}
\affiliation{Laboratoire Lagrange, Universit\'e C\^ote d’Azur, Observatoire de la C\^ote d’Azur, CNRS, Parc Valrose, 06104 Nice Cedex 2, France}

\author[0000-0001-7271-7340]{Julie Hlavacek-Larrondo}
\affiliation{D\'epartement de Physique, Universit\'e de Montr\'eal, Succ. Centre-Ville, Montr\'eal, u\'ebec, H3C 3J7, Canada}

\author[0000-0002-6741-8298]{R. Pierre Martin}
\affiliation{Department of Physics and Astronomy, University of Hawaii at Hilo, 200 W. Kawili St., Hilo, 96720, USA}

\author[0000-0002-3247-5321]{Kathryn Grasha}
\affiliation{Research School of Astronomy and Astrophysics, Australian National University, Weston Creek, ACT 2611, Australia}
\affiliation{ARC Centre of Excellence for All Sky Astrophysics in 3 Dimensions (ASTRO 3D), Australia}

\author[0000-0003-0842-8688]{Natalia Vale Asari}
\affiliation{Departamento de F\'{\i}sica--CFM, Universidade Federal de Santa Catarina, C.P.\ 476, 88040-900, Florian\'opolis, SC, Brazil}
\affiliation{School of Physics and Astronomy, University of St Andrews, North Haugh, St Andrews KY16 9SS, UK}
\affiliation{Royal Society--Newton Advanced Fellowship }

\author[0000-0002-5837-8618]{Théophile Bégin}
\affiliation{D\'epartement de Physique, Universit\'e de Montr\'eal, Succ. Centre-Ville, Montr\'eal, Qu\'ebec, H3C 3J7, Canada}

\author[0000-0002-2478-5119]{Benjamin Vigneron}
\affiliation{D\'epartement de Physique, Universit\'e de Montr\'eal, Succ. Centre-Ville, Montr\'eal, Qu\'ebec, H3C 3J7, Canada}

\author[0000-0002-2457-3431]{Myriam Prasow-Émond}
\affiliation{D\'epartement de Physique, Universit\'e de Montr\'eal, Succ. Centre-Ville, Montr\'eal, Qu\'ebec, H3C 3J7, Canada}

\begin{abstract}
In the first two papers of this series (\citealt{rhea_machine_2020}; \citealt{rhea_machine_2021}), we demonstrated the dynamism of machine learning applied to optical spectral analysis by using neural networks to extract kinematic parameters and emission-line ratios directly from the spectra observed by the SITELLE instrument located at the Canada-France-Hawai'i Telescope. In this third installment, we develop a framework using a convolutional neural network trained on synthetic spectra to determine the number of line-of-sight components present in the SN3 filter (656--683nm) spectral range of SITELLE. We compare this methodology to standard practice using Bayesian Inference. Our results demonstrate that a neural network approach returns more accurate results and uses less computational resources over a range of spectral resolutions. Furthermore, we apply the network to SITELLE observations of the merging galaxy system NGC2207/IC2163. We find that the closest interacting sector and the central regions of the galaxies are best characterized by two line-of-sight components while the outskirts and spiral arms are well-constrained by a single component. Determining the number of resolvable components is crucial in disentangling different galactic components in merging systems and properly extracting their respective kinematics.
\end{abstract}

\section{Introduction}\label{sec:intro}
Systems of merging galaxies are critical tracers of the evolutionary history of the universe according to the well accepted hierarchical model of galaxy formation (e.g. \citealt{conselice_evolution_2014}).
Simulations and observations of interacting systems and galaxy clusters reveal that the merging process is responsible for the creation of a significant fraction of stars as well as modifying the distribution of gas in galaxies and in the intergalactic medium of galaxy clusters (e.g. \citealt{liu_towards_2016}; \citealt{barnes_formation_2002}; \citealt{struck_star_2007}); the mergers are additionally responsible for starbursts occuring after the merger (known as post-merger starbursts) which are phases of intense star formation  (e.g. \citealt{wild_post-starburst_2009}; \citealt{trouille_post-starburst_2013}). The rate and efficiency of the star formation activity vary in time all throughout the merger process and depends on a wide variety of collision parameters but is most importantly related to the galaxies' initial gas content (e.g. \citealt{barnes_transformations_1996}; \citealt{darg_galaxy_2010}; \citealt{donzelli_spectroscopic_2000}).

The evolution and kinematics of star forming gas in \carter{merging systems} can be observed in the optical and near infrared wavelengths. Specifically, several optical lines are used as tracers of ionized gas associated with star-forming regions and other emission line objects:  H\,$\alpha\lambda$6563, and H\,$\beta\lambda$4861, ionized oxygen (i.e. \oii{}$\lambda\lambda$3726,\,3729, and \oiii{}$\lambda\lambda$4959,\,5007), ionized nitrogen (i.e. \nii{}$\lambda\lambda$6548,\,6583), and ionized sulfur (i.e. \sii{}$\lambda\lambda$6716,\,6731) (e.g. \citealt{osterbrock_astrophysics_1989}; \citealt{shields_extragalactic_1990}; \citealt{veilleux_spectral_1987}; the lines' wavelengths are quoted in Angstroms). These strong emission lines help identify excitation mechanisms and chemical abundances present in the gas (\citealt{baldwin_classification_1981}; \citealt{crawford_rosat_1999}; \citealt{kewley_optical_2001}).

To measure the emission lines' flux within a spectrum, it is essential to fit the lines using the proper model with parameters that allow to represent the intensities in all spectral elements. Estimating the number of components of each underlying emission lines (i.e. multiple lines with multiple velocity components as would be expected from merging galaxy systems) is also crucial to extract meaningful information from the lines. Standard methodologies require fitting both single and double component models and computing their Bayes factor, or approximating it using a proxy such as the Akaike (or Bayesian) Information Criterion (AIC; \citealt{rhea_machine_2020}). However, these methods are highly reliant on the accuracy of the fits (e.g. \citealt{kieseppa_akaike_1997}; \citealt{pooley_bayesian_2018}). Several promising new methods have been proposed utilizing machine learning algorithms to solve this problem (\citealt{hampton_using_2017}; \citealt{keown_clover_2019}; \citealt{rhea_novel_2020}). In particular, CLOVER, developed in \cite{keown_clover_2019}, uses a convolutional neural network to classify high-resolution radio emission lines as having either single or double underlying components. In this paper, we expand upon this methodology for medium resolution, Fourier Transform IFU spectra taken by the SITELLE (\textit{Spectromètre Imageur à Transformée de Fourier pour l'Etude en Long et en Large de raies d'Emission}) instrument at the Canada-France-Hawai'i Telescope (\citealt{baril_commissioning_2016}). Each SITELLE observation contains approximately 4 million pixels and, thus, produces 4 million spectra of a given resolution set by the principal investigator ($1 < R < 10,000$; \citealt{drissen_imaging_2014}; \citealt{drissen_sitelle_2019}; \citealt{martin_calibrations_2017}). The instrumental line function of SITELLE is described as a sinc model convolved with a Gaussian to represent intrinsic line broadening
which requires special care during the fitting process (\citealt{martin_optimal_2016}). 
Therefore, the development of machine learning applications for SITELLE demands special treatment of the underlying emission profiles.

In this paper, we explore using a convolutional neural network to classify extragalactic nebulae imaged with SITELLE as having either a single underlying velocity component or two. In $\S$ \ref{sec:meth}, we develop a synthetic set of data, explain the conventional methods for determining the number of underlying components, and develop a convolutional neural network architecture. In $\S$ \ref{sec:res}, we discuss the results of classifying the test set data using the AIC, Bayesian inference, and our convolutional neural network (CNN). We compare the three methodologies, discuss applicability to other instruments, and test out the algorithm on real SITELLE observations of the merging galaxy system NGC2207/IC2163 in $\S$\ref{sec:disc}. Concluding remarks are made in $\S$\ref{sec:conc}. 

\section{Methodology} \label{sec:meth}

In the first two articles of the series, \citet{rhea_machine_2020} $\&$ \citet{rhea_machine_2021}, we explored the application of convolutional and artificial neural networks to calculate the kinematic parameters and flux ratios from SITELLE spectra. In this paper, we employ a convolutional neural network to determine the appropriate number of underlying emission components in \sii$\lambda6717$, \sii$\lambda6731$, \nii$\lambda6584$, H$\alpha$(6563\AA), and \nii$\lambda6548$. To train the algorithm, we start by developing a set of synthetic SITELLE spectra. 

\subsection{Synthetic Data}\label{sec:syn}
The synthetic data is created using the \texttt{ORB} -- \textit{Outil de Reduction Binoculaire} -- software package described in \citet{martin_optimal_2016} and \citet{martin_orbs_2012}. The \texttt{create$\_$cm1$\_$lines$\_$model} is used to create the following emission lines: \sii$\lambda6717$, \sii$\lambda6731$, \nii$\lambda6584$, H$\alpha$(6563\AA), and \nii$\lambda6548$. We use these lines since they are all present in the SN3 (651-685 nm) filter. \texttt{ORB} creates synthetic spectra by taking the spectral resolution, line amplitude, line function, line broadening, and velocity as user inputs. As in other papers, these spectra are representative of those expected from the SIGNALS (\textit{Star formation, Ionized Gas, and Nebular Abundances Legacy Survey}; \citealp{rousseau-nepton_signals_2019}). We randomly sample the resolution from a uniform distribution varying between R=4800 and R=5000; this variation mimics expected resolution variations within the SITELLE field of view, that are observed in SITELLE data cubes. 

The velocity value is randomly chosen from a uniform distribution between $-500$~\kms{} and $500$~\kms; these values well encompass the typical rotation velocity range for galaxies found in the SIGNALS program. Additionally, we select the broadening from an uniform distribution between 5 \kms{} and 20 \kms{}. We selected the lower limit since SITELLE cannot resolve broadening beneath this level (\citealt{rousseau-nepton_signals_2019}). We note that shocks in the ISM are not being considered here -- they would require an increase in the maximum broadening. We verified that the chosen lower limit does not impact the classification process by testing different lower bounds (3 \kms{} and 1 \kms{}).
We selected 20 \kms{} since it is the expected upper limit of dispersion in HII regions based onon previous experience in SITELLE analysis (\citealt{rhea_machine_2020}).
Moreover, we randomly vary the signal-to-noise ratio concerning the H$\alpha$ emission between 5 and 30. We propagate the noise to the other emission lines by adding it to each spectral channel. The noise factor itself is sampled from Gaussian with a sigma of 1. Therefore, each spectrum has a different signal-to-noise ratio described by a Gaussian centered between 5 and 30 with a sigma of 1.
The last feature required to create synthetic spectra is the relative amplitude of the strong emission lines. Following the methodology described in detail in \citet{rhea_machine_2020},  relative  line  amplitude  of  the  HII  regions  were  sampled  from  the  Mexican  Million  Models database (3MdB; \citealt{morisset_virtual_2015}) BOND simulations (\citealt{asari_bond_2016}).

We created a set of 10,000 spectra with a single component; therefore, there are five emission lines present (\sii$\lambda6717$, \sii$\lambda6731$, \nii$\lambda6584$, H$\alpha$, and \nii$\lambda6548$) with the same velocity and broadening values. An additional 10,000 spectra with two components was also created; therefore, there are two sets of five emission lines present with the same velocity and broadening values within each set. Each set has a different velocity and broadening value. The amplitudes for each set come from different instances of the \texttt{BOND} simulations from \texttt{3MdB}. Thus, we have 20,000 spectra in total.
Following standard procedure, we use 70\% of the synthetic data for the training set, 20\% for the validation set, and 10\% for the test set (e.g., \citealt{breiman_random_2001}). We note that individual lines may have different kinematics; however, we do not consider this in our analysis for simplicity's sake.

\subsection{Bayesian Inference}\label{sec:bayes}
In the past decade, Bayesian statistics have become house-hold tools for astronomers; Bayesian techniques are used to study eclipsing exoplanet signals (\citealt{taaki_bayesian_2020}; \citealt{ruffio_bayesian_2018}), fitting spectra in order to extract model parameters (\citealt{sereno_bayesian_2016}; \citealt{sharma_markov_2017}), and model comparison (\citealt{jenkins_power_2011}; \citealt{trotta_applications_2007}).
In this section, we outline the mathematics behind Bayesian inference, how it can be used to compare models, and its practical implementations in \texttt{python}.

Bayesian inference is the process of uncovering underlying parameter distributions through the use of Bayes' theorem. Bayes' theorem states 
\begin{equation}\label{eqn:bayes}
    P(\Theta_M|Y,I) = \frac{P(Y|\Theta_M,I)\times P(\Theta_M|I)}{P(Y|I)},
\end{equation}
where $\Theta_M$ describes a set of model parameters for model $M$, $Y$ is the set of observed data, and $I$ represents any assumed information. The left-hand side of the equation is known as the posterior distribution, and it describes the probability distribution of model parameters given the existing data. 
On the right-hand side, we have the likelihood distribution, $P(Y|\Theta_M,I)$, which describes the distribution of the data given a set of model parameter values, the prior distribution, $P(\Theta_M,I)$, describing our prior knowledge or believes regarding the distribution of the model parameters, and the evidence, $P(Y|I)$ or $Z$. In many scenarios, Bayes' theorem is rewritten as 
\begin{equation}
    P(\Theta_M|Y,I) \propto P(Y|\Theta_M,I)\times P(\Theta_M|I),
\end{equation}
where the evidence is ignored. However, in the case of model comparison, we wish to calculate the Bayes factor, which is defined as the ratio of the evidences calculated using two distinct models. Therefore, the evidence must be calculated; note that the evidence is defined as the integral over all the likelihood multiplied by the prior model parameters.
\begin{equation}
    Z = \int_{\theta_M \in \Theta_M}P(Y|\Theta_M,I)\times P(\Theta_M|I) d\theta_m.
\end{equation}

We write the Bayes' factor, $P(M|D)$, as 
\begin{equation}
    P(M|D) = \frac{Z_1}{Z_2}
\end{equation}

We adopt a standard Gaussian likelihood function of the following form:
\begin{equation}
    P(Y|M_\Theta,I) = \frac{1}{\sqrt{2\pi}\sigma_0}\exp{\Big(\frac{-(Y-Y_M(M_\Theta))^2}{2\sigma_0^2}\Big)}
\end{equation}
where $Y_M(M_\Theta)$ is the assumed model evaluated given the model parameters $M_\Theta$ and $\sigma_0$ are the measurement errors. For the sake of the calculations simplicity (discussed in \S \ref{sec:comparison}), we employ a standard Gaussian to model the emission line. The Gaussian has the following form:
\begin{equation}
    Y_M(M_\Theta) = Y_M(A, x_0, \sigma) = A \exp{\Big(\frac{-(x-x_0)^2}{2\sigma^2}\Big)}
\end{equation}
where $A$ is the amplitude of the Gaussian, $x$ is the wavelength channel, $x_0$ is the position of the line, and $\sigma$ is the width of the line. We emphasize that we calculate the posteriors for the parameters $M_\Theta = [A, x_0, \sigma]$ in the case of a single Gaussian model. We adopt linearly uniform priors for all the fit parameters.

Unfortunately, the calculation of the evidence is exceptionally costly in high-dimensional spaces, and alternative methods to calculate the evidence are required. 
Nested sampling is one of the most popular methods for accurately and rapidly calculating the evidence (e.g. \citealt{skilling_nested_2006}; \citealt{chopin_properties_2010}). We, therefore, use the nested sampling code \texttt{dynesty} for our calculations (\citealt{speagle_dynesty_2020}). Since the calculations are computationally expensive, we explore a faster algorithm -- a convolutional neural network.

\subsection{Convolutional Neural Networks} \label{sec:cnn}
Neural networks, and their extension convolutional neural networks, have been used extensively to solve astrophysical problems (e.g. \citealt{baron_machine_2019}; \citealt{davies_using_2019}; \citealt{aniyan_classifying_2017}; \citealt{kim_stargalaxy_2017}). Recently, convolutional neural networks have been shown to be efficient at rapidly and accurately extracting parameters from spectra (e.g. \citealt{fabbro_application_2018}; \citealt{obriain_cycle-starnet_2021};\citealt{keown_clover_2019}; \citealt{rhea_machine_2020}). For a detailed review of convolutional neural networks, we direct the reader towards the following works: \cite{kuo_understanding_2016}, \cite{liu_towards_2016}, \cite{khan_survey_2020}. In particular, by interpreting model selection as a classification problem, neural networks with a softmax activation on the last layer can be used as a substitute to the reference Bayesian model selection approach, with a substantial gain in computational complexity.

For this work, we adopt the structure of the network STARNET developed in \cite{fabbro_application_2018} and used in our previous work \citep{rhea_machine_2020}. We note that the hyperparameters (i.e. filters) are tuned separately and are thus different from those used in \citet{fabbro_application_2018}. The network is as follows:
\begin{enumerate}
    \item Convolutional layer with 4 filters of 16 elements activated with the \texttt{relu} function
    \item Convolutional layer with 4 filters of 8 elements activated with the \texttt{relu} function
    \item Pooling layer with a size of 8 elements  
    \item Flattening Layer
    \item 20\% Dropout
    \item Fully connected layer with 1000 nodes activated with the \texttt{relu} function and regularized using $\ell_2$ regularization 
    \item 20\% Dropout
    \item Fully connected layer with 1000 nodes activated with the \texttt{relu} function and regularized using $\ell_2$ regularization 
    \item Fully connected output layer with two nodes activated with the \texttt{softmax} function
\end{enumerate}
We note that the use of the \texttt{softmax} activation function in the output layer not only generates a binary classification but also assigns a probability of accuracy to the classification.

Although we adopt the convolutional neural network, STARNET, described in detail in \citealt{fabbro_application_2018}, we apply a complete suite of hyper-parameter tuning. Using the \texttt{sklearn} grid search implementation, we optimized the number of convolutional kernels in each layer (4 in both layers) and the length of each filter (16 elements in the first layer and 8 elements in the second layer. We also optimized the number of spectra fed into the network at a time (i.e., the optimal batch size; we found this to be 2 spectra), the best optimization algorithm, \texttt{Adam}, the length of the convolutional pooling in each layer, 8.
These results are summarized in table \ref{tab:hyper}.

\begin{table}
    \label{tab:hyper}
    \begin{tabular}{|c|c|}
    \hline 
        Hyper Parameter & Value \\ \hline \hline 
        Batch Number & $2$ \\ \hline
        $\ell_2$ Regularization & $5e^{-5}$\\ \hline
        Optimizer & Adam \\ \hline 
        Kernel Initializer & Glorot Uniform \\ \hline 
        Max Epochs & $25$\\ \hline 
    \end{tabular}
\end{table}

Since the resolution is free to vary uniformly from 4800 to 5000 and the network requires inputs of a constant size, we interpolate the spectra using linear cardinal splines using a grid taken from a reference spectrum with a resolution set to 5000. This does not affect the fidelity of the spectrum.

\section{Results} \label{sec:res}
In this section, we discuss the results of the Bayesian inference and CNN algorithms. We use confusion matrices to compare the efficiencies of the different methodologies. Confusion matrices compare the true categorizations (Y-axis) with the predicted categorizations (X-axis). A perfectly diagonal confusion matrix indicates that the method accurately predicts the number of line-of-sight components 100\% of the time. Cross terms represent misclassifications. The values quoted are percentages.

\subsection{Bayesian Inference}
\begin{figure}
    \caption{Confusion matrix for the Bayesian inference calculations.}
    \label{fig:bi_conf}
    \centering
    \includegraphics[width=0.45\textwidth]{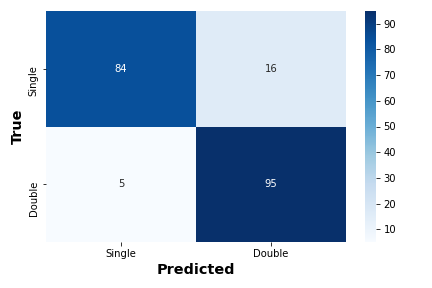}
\end{figure}

Following the procedure described in $\S$\ref{sec:bayes}, we fit a single Gaussian model and a double Gaussian model using the \texttt{dynesty} implementation of static nested sampling to the same test set used in the CNN calculations. 
We fit only the H$\alpha$ line in order to reduce the computational cost (see \S \ref{sec:comparison} for a more detailed discussion on the repercussions of this decision).
We then used the reported outputs to calculate the Bayes' factor. The double component model is considered favorable if the Bayes' factor is greater than 1.08 (this value was experimentally determined to return the best classifications\footnote{In order to determine the cutoff Bayes factor we varied the cutoff from 1 to 5 in increments of 0.05 and calculated the f1-score given the chosen factor. We then took the Bayes factor that returned the best f1-score.
}); otherwise, a single component model is favorable. We used the static nested sampling implementation over the more accurate dynamic nested sampling implementation since the latter increased the computational costs by more than an order of magnitude -- this is further discussed in $\S$\ref{sec:disc}. The results are shown in figure $\ref{fig:bi_conf}$. We investigated the cases in which a double component model was mistaken for a single component model (approximately 15\% of the time) and discovered that the majority of misclassifications occurred when the absolute velocity difference between the two components was less than 250 \kms{} (see the next subsection for further discussion). 

\subsection{Convolutional Neural Network}
\begin{figure}[!h]
    \centering
    \includegraphics[width=0.45\textwidth]{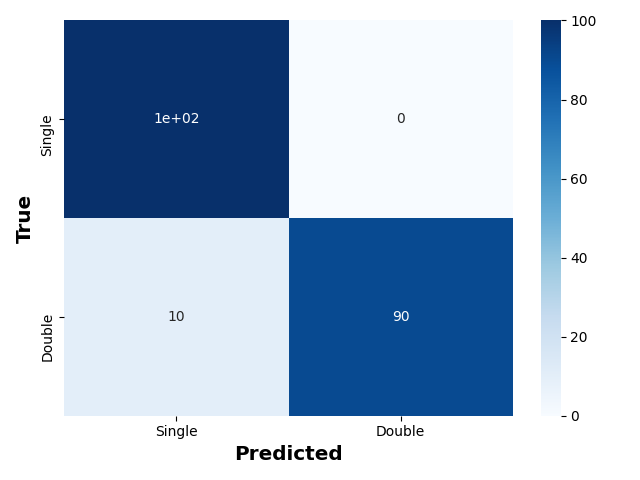}
    \caption{Confusion matrix for the convolutional neural network described in $\S$\ref{sec:cnn}.}
    \label{fig:cnn_conf}
\end{figure}

The convolutional neural network was trained and validated using 70\% and 20\% of the synthetic data. At the end of the process, the overall accuracy of the algorithm was 93\%. The mse (mean squared error) on both the test and validation set is 0.44; this indicates that the network is not over-fitting. Over-fitting occurs when the network learns the test set too well and cannot generalize to other data (i.e., the validation or test set). The confusion matrix for the test set (see figure \ref{fig:cnn_conf}) reveals near-perfect precision in classifying single line-of-sight component spectra. Moreover, the matrix demonstrates that the network classifies double line-of-sight component spectra accurately 90\% of the time; the remaining 10\% are misclassified as single components.  We note that this is less than the 95\% accuracy of the Bayesian methodology.

In order to establish that the network is not over-fitting, we applied the standard k-fold cross-validation with $k=10$. Additionally, we applied a modified $k$-fold cross-validation algorithm in which only the training and validation set are varied while the test set remains consistent across all folds. The reported accuracy of the model for each fold regardless of implementation remained constant (less than 5\% variation). This is a strong indication that the network is not over-fitting, and, thus, it can be generalized to unseen data.

\begin{figure}[!h]
    \centering
    \includegraphics[width=0.45\textwidth]{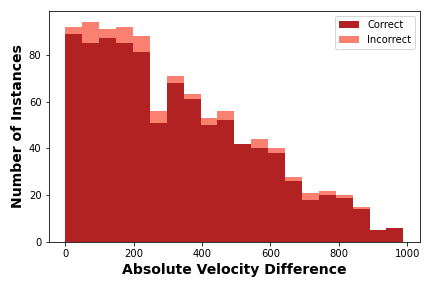}
    \caption{Absolute velocity difference vs number of instances for mis-categorized spectra containing two underlying components.}
    \label{fig:vel_diff}
\end{figure}

Moreover, we examined the regimes in which the network fails to predict the number of underlying components accurately. In figure \ref{fig:vel_diff}, we plot a histogram of the absolute velocity difference for the test set spectra with double line-of-sight components. The stacked histogram shows that the incorrectly categorized spectra (rose) are primarily clustered around absolute velocity differences less than 250 \kms{}. This trend is expected since, at these low absolute velocity differences, the components are blended considerably. We show four spectra illustrating this issue in appendix \ref{app:spectra}.

\subsection{Other Machine Learning Algorithms}\label{sec:MLA}
\begin{figure}[!h]
    \centering
    \includegraphics[width=0.45\textwidth]{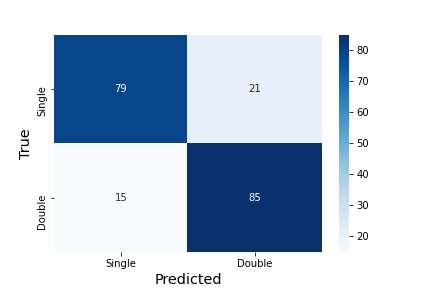}
    \caption{Confusion matrix for the random forest classifier tested in this work}
    \label{fig:RF-confmat}
\end{figure}

In addition to using a convolutional neural network to estimate the number of emission-line components, we tested another common machine learning algorithm for binary classification problems: the random forest. While random forests are themselves complex machine learning algorithms, we do not discuss them in detail here since the purpose of this section is to demonstrate the efficacy of other machine learning algorithms to solve our problem; instead we point the reader to (\citealt{breiman_random_2001}) and the references therein. Although we test several configurations for the random forest classifier, our final classifier has a total of 50 independent estimators (decision trees) with a maximum recursion depth of 5 levels. The split criterion is based on entropy calculations instead of gini calculations. As evidenced by figure \ref{fig:RF-confmat}, the random forest algorithm does not achieve the same fidelity in categorization as the neural network.

\section{Discussion}\label{sec:disc}
\subsection{Comparison of Methods}\label{sec:comparison}
As evidenced by the confusion matrices shown in figures \ref{fig:bi_conf} and \ref{fig:cnn_conf}, the convolutional neural network outperforms the Bayesian inference classifications in the single component case. Notably, the CNN slightly (3\%) under-performs the Bayesian Inference analysis for the classification of double line-of-sight components.
Since the two methods were tested on the same set of data (the CNN's test set), the methodologies' comparison is unbiased with regards to the data they are tested on. Although the Bayesian inference methodology returns similar results to that of the CNN, the method takes two orders of magnitude more computational time, rendering it impractical. The Bayesian method takes approximately 200 hours to fit the test set, while the network takes 45 minutes to train and approximately 1 second to predict the test set. Additionally, the creation of the synthetic data set takes approximately 20 minutes. Therefore, the total time for the CNN is slightly over 1 hour.

Generally, Bayesian Inference sets a baseline for fitting results; however, our results indicate that the CNN is outperforming the Bayesian inference model estimates for the single line-of-sight component scenario since it is more accurate for single line-of-sight component classifications. We attribute this discrepancy to the chosen calculation of the Bayesian evidence and the model used. In order to reduce the number of parameters to a more computationally manageable size, we only fit a Gaussian (or two Gaussians) to the H$\alpha$ component. In doing so, we must marginalize over a 3 or 6 dimensional space, respectively (there are 3 components used to describe each Gaussian in the Bayesian approach). If  we were to fit all five lines, we would be required to marginalize over a 15 or 30 dimensional space thus further increasing the required computational time. Therefore, we emphasize that the CNN is outperforming the Bayesian approach only because the Bayesian approach is incomplete since it does not fully treat the ILS of SITELLE nor does it take into account every line present in the SN3 spectra. We further note that a Bayesian approach taking into account all lines and the proper ILS is not implemented since it is too computationally expensive at this time.

We also note the work of \cite{gonzalez-gaitan_spatial_2019} in which the authors use spatial information and prior estimations on the spatial correlations to drastically reduce the required computational time using Bayesian methods. However, a direct comparison between our works is beyond the scope of this paper.

\subsection{Differing Resolution}
\begin{figure}[!hbt]
    \centering
    \includegraphics[width=0.48\textwidth]{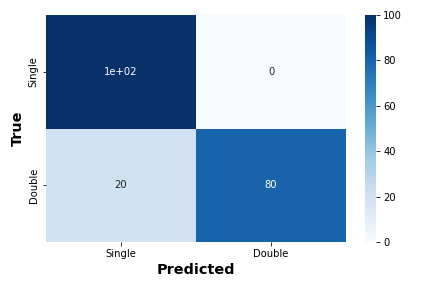}
    \caption{Confusion matrix for the convolutional neural network applied to spectra with $R\sim3000$. The rows sum to 100 (unless there is a discrepancy due to rounding).}
    \label{fig:conf-3000}
\end{figure}

In order to port this methodology to observations that are not part of the SIGNALS collaboration, we create an alternative test set at a different resolution. We select a resolution of 3000 to test the algorithm's efficacy on lower-resolution spectra. After creating a set of synthetic data following the same procedure described in section $\S$\ref{sec:syn} with $R\sim3000$, we train, validate, and test the algorithm on this lower-resolution set. Figure \ref{fig:conf-3000} demonstrates that the categorizations' fidelity does not change for single line-of-sight component spectra. At the same time, the accuracy is reduced by 10\% for double line-of-sight component spectra. Additionally, we followed the same procedure for $R\sim1000$ spectra; however, the network fails to achieve the same level of efficacy due to the shallow resolution. We, therefore, do not suggest using the network on spectra with a resolution considerably lower than $R\sim3000$. 

\subsection{Application to More Components}\label{sec:three-comp}
\begin{figure}
    \centering
    \includegraphics[width=0.48\textwidth]{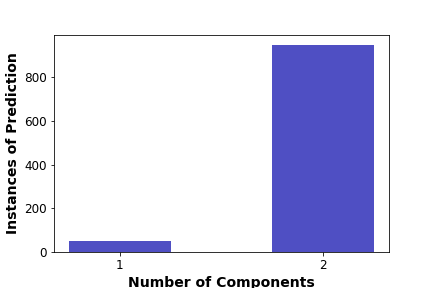}
    \caption{Histogram portraying the number of instances the network classified a spectrum containing three underlying emission components as only have one or two components. There were a total of 51 spectra classified as having only a single component while the remaining 949 spectra were classified as having two spectra.}
    \label{fig:three-comp}
\end{figure}

Although we have only considered the problem as a binary classification (i.e. either single or double component) due to the constraints of the SIGNALS observing campaign, it is possible that a spectrum can contain more than two components. Therefore, we constructed a set of 1,000 synthetic spectra comprised of three underlying emission components following the same methodology as outlined in $\S$\ref{sec:syn}. We then apply our network to the spectra. Figure \ref{fig:three-comp} demonstrates that the network classifies the majority ($\sim$95\%) of the three-component spectra as having two components. Therefore, we suggest the following interpretation if a region is suspected to have more than two components: a classification as two components should be considered as \textit{at least} two components. These results are similar to those found in \citealp{rhea_novel_2020}.  

\subsection{Application to Real SITELLE Observations}
\begin{figure*}[!thb]
    \centering
    \includegraphics[width=0.9\textwidth]{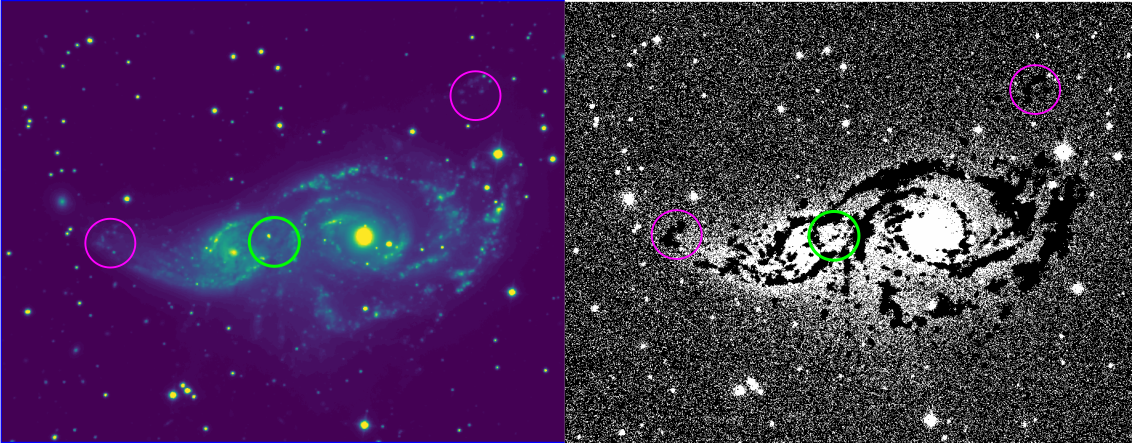}
    \caption{ Left: Deep SITELLE image of the N2207/IC2163 system created with \texttt{ORCS}. This panel shows the stacked optical emission in the component galaxies. The colors (from blue to yellow) indicate the integrated flux in a spaxel. Several structures such as the spiral arms, bulges, tidal tails, and diffuse emission regions stand out. Right: Component map for the NGC2207/IC2163 system. White pixels correspond to double component emission. Black pixels correspond to single component emission. The green circle highlights the mixing region while the magenta circles highlight diffuse regions.}
    \label{fig:comp-deep}
\end{figure*}

In order to demonstrate the utility of this methodology, it is applied to actual SITELLE observations at R$\sim$3000 of the interacting pair of spiral galaxies NGC 2207 and IC 2163 (P.I. R. Pierre Martin). This system (D = 35 Mpc) is undergoing a grazing collision and has been studied from X-rays to cm wavelengths (e.g. \citealt{kaufman_ngc_2012}; \citealt{elmegreen_alma_2017}). The closest interaction of both galaxies is estimated to have happened 200--400 Myr ago (\citealt{kaufman_ngc_2012}) and, according to models, both galaxies are expected to merge in the future (\citealt{struck_star_2007}). SITELLE observations cover the entire interaction, including zones not severely affected by the collision, tidal tails, and the zone of closest proximity between both galaxies. Therefore, this target presents an ideal test-bed for our algorithm. We note that no $R\sim5000$ SITELLE observations of merging galaxies exist at the time of writing this paper.

In addition to applying the neural network to the SITELLE data cube, we obtain a deep image of the SITELLE cube using \texttt{ORCS}. The deep image is a 2d, stacked representation of the 3d data cube (i.e., the spectral information is compressed to a single pixel). The neural network returns a map of the galaxy where each pixel is designated as having either a single or double component. A comparison of these maps (the component map and the deep image) is shown in figure \ref{fig:comp-deep}. The component map reveals several interesting features: double components in the bulges of each spiral galaxy, the double component emission in the current closest proximity zone (green circle), and single component emission in the spiral arms. Additionally, the component map classifies diffuse regions (indicated in magenta circles) as single component emission regions. We note that the HII regions in the center of each galaxy are also categorized by double component emission. These findings are consistent with the literature (i.e. \citealt{soto_emission-line_2012}). 
Since we use a \texttt{softmax} activation function in the final layer of the network, we also obtain probability maps. These maps in general can be used to make probability cuts reflecting regions for which the network is sure of its classification. We discuss this in more detail in the appendix $\ref{app:probs}$. \carter{We hypothesize that since the bulges are being classified with a high probability as having two components these regions have at least two components; this is consistent with our findings in $\S$\ref{fig:three-comp}. These results indicate that the network's classifications should be accepted cautiously for regions which are heavily contaminated by astrophysical processes such as AGN or a bright stellar component.}
In regions that correspond to the deep image background (i.e., not part of either galaxy in the system), the component map is noisy. This indicates that the component map is reliable only for regions of sufficient flux -- which is expected.
We stress that we do not expect to be able to resolve more than two emission-line components given the resolution of SIGNALS SN3 observations (R$\sim$5000) and the distance to the objects. Expanding the network to be a multi-component classifier is a viable avenue for future research.

\subsection{Scientific Implications}
Determining the number of resolvable line-of-sight components in galaxy spectra carries several important scientific implications. In situations where two line of sight components are present and resolvable, it is important to fit both lines separately instead of treating the emission as a single component for several reasons. Distinguishing a region of multiple component line-of-sight emission from a large blended region will allow for the correct characterization of thermal broadening in \hii{} regions which will help untangle the reaction of the gas to the ionizing source. Moreover, if the two components have considerably distinct kinematic parameters (velocity and broadening), treating the emission as a single component will lead to a misclassification of the kinematics of the region which can lead to incorrect interpretations of the star forming mechanisms present. In effect, a misunderstanding of the number of resolvable line-of-sight components present will lead to a misrepresentation of the underlying physics. 

Additionally, mapping out regions in which two line-of-sight components is crucial in disentangling galaxy systems undergoing a merger. In doing so, the kinematics of the component galaxies can be studied; moreover, treating each galaxy's emission separately leads to more accurate calculations of the emission line fluxes. This is crucial for understanding the underlying ionization mechanisms at play in mixing regions of merger systems (e.g., \citealt{baldwin_classification_1981}; \citealt{kewley_optical_2001}; \citealt{kewley_understanding_2019}; \citealt{rich_galaxy_2015}). Furthermore, by applying the networks developed in \cite{rhea_machine_2020} and \cite{rhea_machine_2021} to the appropriate regions, kinematic parameters and emission-line ratios can be recovered. The authors note that those networks function only for a single component spectrum; future work will demonstrate their extension to double component spectra.

\subsection{Prospects of application in other instruments}
Although we focus on SITELLE data cubes in this article, the methodology described herein can be readily ported to any other high-resolution integral field unit that has access to the \nii{} doublet, the \sii{} doublet, and H$\alpha$ emission lines such as the Multi Unit Spectroscopic Explorer, MUSE (e.g., Bacon et al. 2010; \citealt{kreckel_mapping_2019}; \citealt{kreckel_measuring_2020}; \citealt{foster_magpi_2020}). When applying this methodology to other instruments, it is essential to repeat the creation of synthetic data and network training, validation, and testing if the instrumental response function is not a \texttt{sincgauss}. 

\section{Conclusions}\label{sec:conc}
Machine learning algorithms present a novel approach for spectral analysis. In the first two papers of this series, we demonstrated the efficacy of convolutional and traditional neural networks at extracting kinematic parameters and emission-line flux ratios from SITELLE spectra (\citealt{rhea_machine_2020}; \citealt{rhea_machine_2021}). In this paper, the third of the series, we develop a convolutional neural network to classify spectra as having either a single or double line-of-sight component. This systematic method will be critical for disentangling components in merger systems, HII regions, and supernova remnants. We demonstrate that the network outperforms Bayesian inference model comparisons in the single-component case and recovers similar accuracy in the double-component case. Moreover, the computational costs associated with the CNN training and subsequent application are several of orders of magnitude lower than the cost of the full Bayesian approach. In order to demonstrate the applicability of this network to real SITELLE data, we apply the network to actual SITELLE observations of the merging galaxy system NGC2207/IC2163. We find that the central regions of the individual galaxies and the closest proximity region are best categorized by double components. At the same time, the spiral arms and diffuse emission on the outskirts are best described by single component emission. The code can be found, along with code from the previous papers, at \url{https://github.com/sitelle-signals/Pamplemousse}.

\acknowledgments
The authors would like to thank the Canada-France-Hawaii Telescope (CFHT) which is operated by the National Research Council (NRC) of Canada, the Institut National des Sciences de l'Univers of the Centre National de la Recherche Scientifique (CNRS) of France, and the University of Hawaii. The observations at the CFHT were performed with care and respect from the summit of Maunakea which is a significant cultural and historic site.

C. R. acknowledges financial support from the physics department of the Universit\'e de Montr\'eal,  IVADO, and the FRQNT.
J. H.-L. acknowledges support from NSERC via the Discovery grant program, as well as the Canada Research Chair program.
NVA acknowledges support of the Royal Society and the Newton Fund via the award of a Royal Society--Newton Advanced Fellowship (grant NAF\textbackslash{}R1\textbackslash{}180403), and of Funda\c{c}\~ao de Amparo \'a Pesquisa e Inova\c{c}\~ao de Santa Catarina (FAPESC) and Conselho Nacional de Desenvolvimento Cient\'{i}fico e Tecnol\'{o}gico (CNPq). We thank the anonymous referee for their insightful comments.

\software{python \citep{van_rossum_python_2009}, numpy \citep{van_der_walt_numpy_2011}, scipy \citep{virtanen_scipy_2020}, matplotlib \citep{hunter_matplotlib_2007}, pandas \citep{mckinney_data_2010}, seaborn \citep{michael_waskom_mwaskomseaborn_2017}, \citep{robitaille_astropy_2013}, tensorflow \citep{abadi_tensorflow_2015}, keras \citep{chollet_keras_2015}}.

\newpage

\appendix 
\section{Illustrative Spectra}\label{app:spectra}
In this section we show four spectra which illlustrate correct classifications and misclassifications for double component spectra. We further divide the spectra by the absolute velocity difference of their components; we use 250 \kms{} as the divider. The top row images (a and b) are spectra with two components that were correctly categorized while the bottom row images (c and d) are spectra with two components that were incorrectly categorized. Similarly, the left images (a and c) contain components with an absolute velocity difference less than 250 \kms{} while the right images (b and d) contain components with an absolute velocity difference greater than 250 \kms{}.

\begin{figure*}
\gridline{\fig{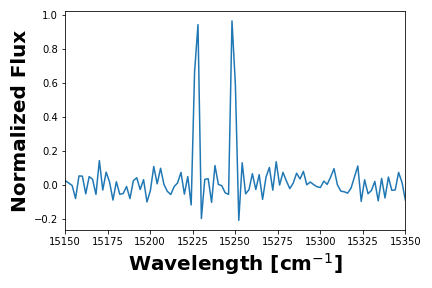}{0.5\textwidth}{(a) Correctly classified double component spectrum with an absolute velocity difference of 85\kms{}}
          \fig{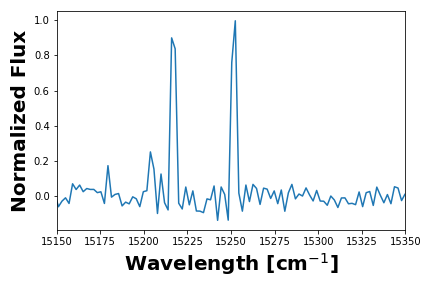}{0.5\textwidth}{(b) Correctly classified double component spectrum with an absolute velocity difference of 500\kms{}}
          }
\gridline{\fig{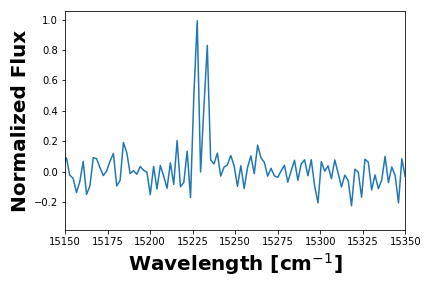}{0.5\textwidth}{(c) Incorrectly classified double component spectrum with an absolute velocity difference of 146\kms{}}
          \fig{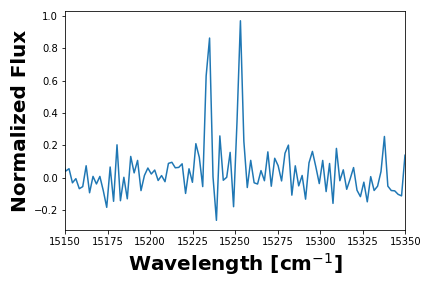}{0.5\textwidth}{(d) Incorrectly classified double component spectrum with an absolute velocity difference of 465\kms{}}
          }          
\caption{}
\label{fig:spectra_plots}
\end{figure*}

\section{Corner Plots from Dynesty}
In this section we show the corner plots for a spectrum with two emission-line components assuming a likelihood with only a single Gaussian and a likelihood with two Gaussians.

\begin{figure}
    \centering
    \includegraphics[width=0.9\textwidth]{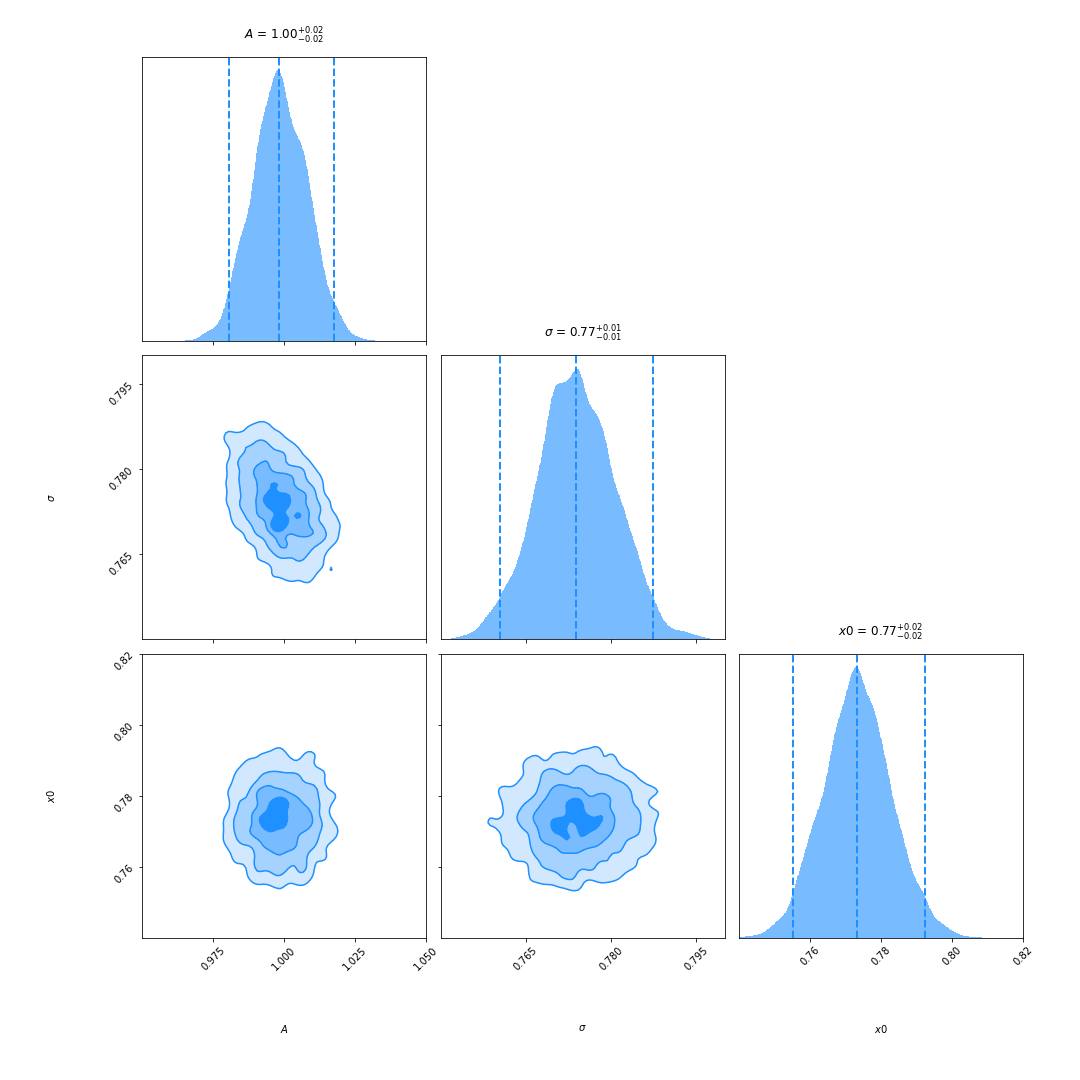}
    \caption{Corner plot from \texttt{dynesty} for the case of a spectrum with two underlying emission-line components modeled with a single Gaussian.}
    \label{fig:double-single}
\end{figure}

\begin{figure}
    \centering
    \includegraphics[width=0.9\textwidth]{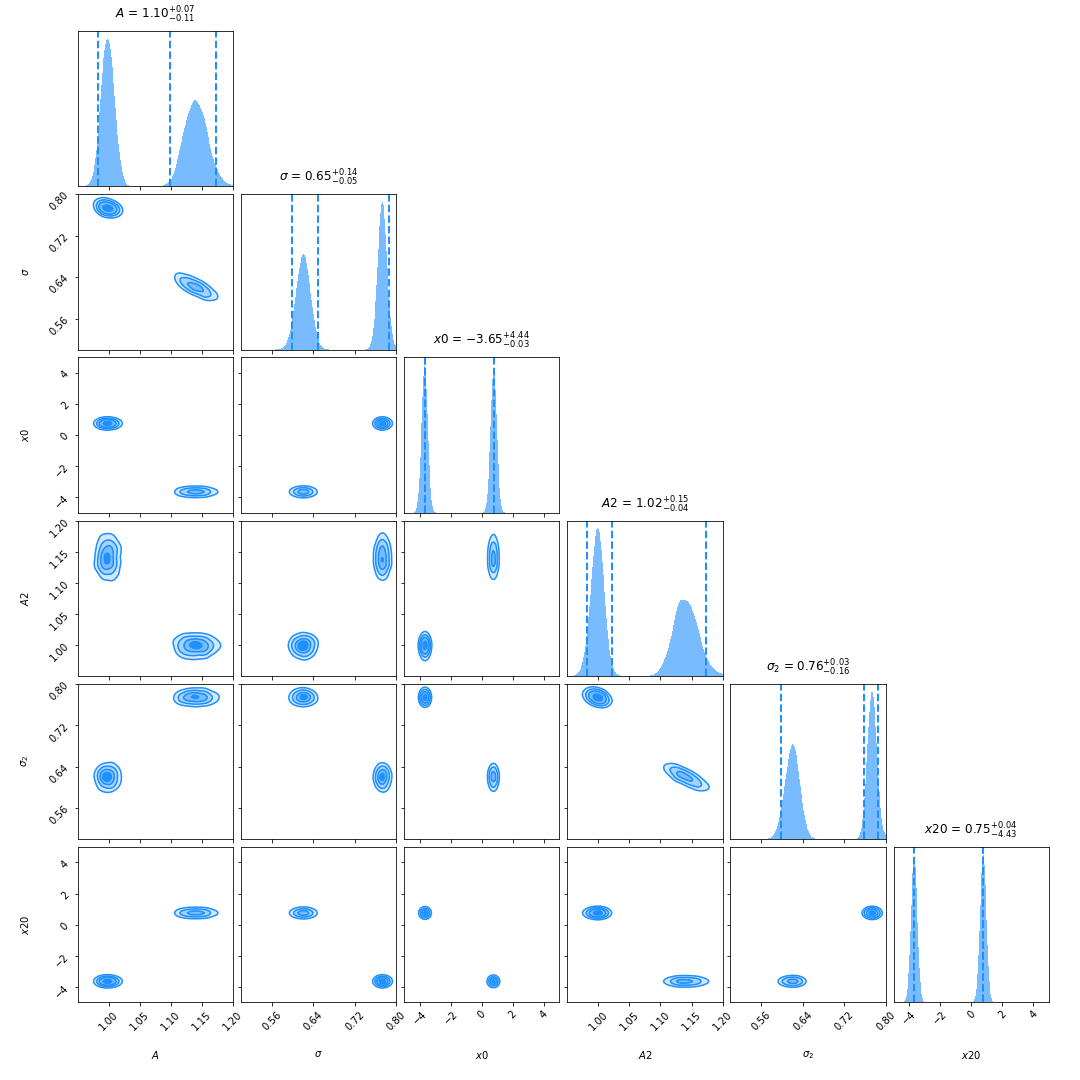}
    \caption{Corner plot from \texttt{dynesty} for the case of a spectrum with two underlying emission-line components modeled with two Gaussians.}
    \label{fig:double-double}
\end{figure}

\section{NGC 2207 Classification Probability Map}\label{app:probs}
\begin{figure}
    \centering
    \includegraphics[width=0.85\textwidth]{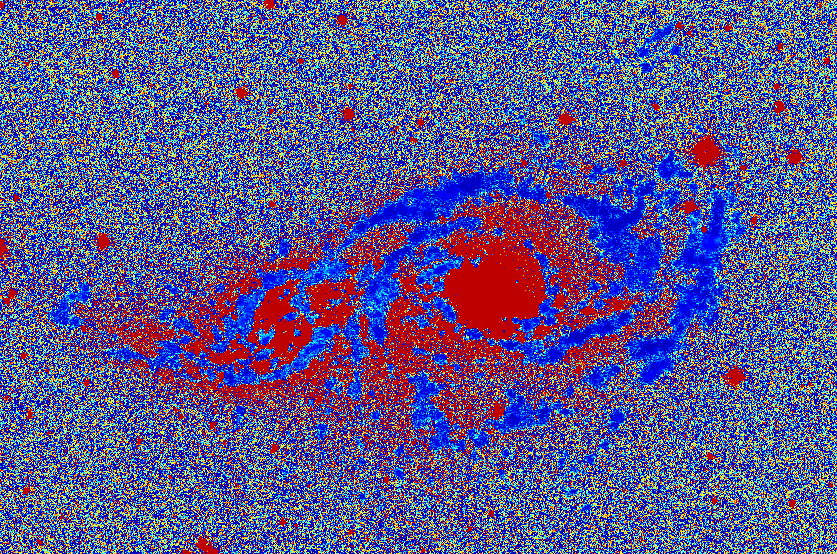}
    \caption{Classification probability map of the NGC 2207 merging system. Red regions correspond to a probability of 1, while regions in dark blue correspond to a probability of approximately 0.85.}
    \label{fig:probs}
\end{figure}

As mentioned, we obtain a probability of the classification being correct along with the classification. Since this is a binary problem, the probabilities are bound between 0.5 and 1.0. As we see in the figure below (\ref{fig:probs}), the regions of lower surface brightness and spiral arms are classified with a probability generally between 0.85 and 0.9; on the contrary, the bulges are classified with 100\% probability. 
\carter{ A classification probability of 100\% is unrealistic; however, we demonstrated in $\S$ \ref{sec:three-comp} that spectra more complicated than having a simple single or double emission component are classified as having two components.
This likely indicates that the bulge regions are extremely complicated and may need more than two emission components to accurately model. Although a detailed analysis of this is beyond the scope of the paper, the collaboration is studying the bulges of galaxies in the process of a merger in more detail.} 
Importantly, the probabilities vary considerably in regions with low signal-to-noise ratios. Therefore, we suggest masking out these areas when using the network's results for subsequent spectral analysis.

\bibliography{SitelleML3}{}

\begin{thebibliography}{}
\expandafter\ifx\csname natexlab\endcsname\relax\def\natexlab#1{#1}\fi
\providecommand{\url}[1]{\href{#1}{#1}}
\providecommand{\dodoi}[1]{doi:~\href{http://doi.org/#1}{\nolinkurl{#1}}}
\providecommand{\doeprint}[1]{\href{http://ascl.net/#1}{\nolinkurl{http://ascl.net/#1}}}
\providecommand{\doarXiv}[1]{\href{https://arxiv.org/abs/#1}{\nolinkurl{https://arxiv.org/abs/#1}}}

\bibitem[{Abadi {et~al.}(2015)Abadi, Agarwal, Barham, Brevdo, Chen, Citro,
  Corrado, Davis, Dean, Devin, Ghemawat, Goodfellow, Harp, Irving, Isard, Jia,
  Jozefowicz, Kaiser, Kudlur, Levenberg, Mane, Monga, Moore, Murray, Olah,
  Schuster, Shlens, Steiner, Sutskever, Talwar, Tucker, Vanhoucke, Vasudevan,
  Viegas, Vinyals, Warden, Wattenberg, Wicke, Yu, \&
  Zheng}]{abadi_tensorflow_2015}
Abadi, M., Agarwal, A., Barham, P., {et~al.} 2015, 19

\bibitem[{Aniyan \& Thorat(2017)}]{aniyan_classifying_2017}
Aniyan, A.~K., \& Thorat, K. 2017, 230, 20, \dodoi{10.3847/1538-4365/aa7333}

\bibitem[{Asari {et~al.}(2016)Asari, Stasińska, Morisset, \&
  Fernandes}]{asari_bond_2016}
Asari, N.~V., Stasińska, G., Morisset, C., \& Fernandes, R.~C. 2016, 460,
  1739, \dodoi{10.1093/mnras/stw971}

\bibitem[{Baldwin {et~al.}(1981)Baldwin, Phillips, \&
  Terlevich}]{baldwin_classification_1981}
Baldwin, J.~A., Phillips, M.~M., \& Terlevich, R. 1981, 93, 5,
  \dodoi{10.1086/130766}

\bibitem[{Baril {et~al.}(2016)Baril, Grandmont, Mandar, Drissen, Martin, \&
  Rousseau-Nepton}]{baril_commissioning_2016}
Baril, M., Grandmont, F., Mandar, J., {et~al.} 2016, 9908

\bibitem[{Barnes(2002)}]{barnes_formation_2002}
Barnes, J.~E. 2002, 333, 481, \dodoi{10.1046/j.1365-8711.2002.05335.x}

\bibitem[{Barnes \& Hernquist(1996)}]{barnes_transformations_1996}
Barnes, J.~E., \& Hernquist, L. 1996, 471, 115, \dodoi{10.1086/177957}

\bibitem[{Baron(2019)}]{baron_machine_2019}
Baron, D. 2019

\bibitem[{Breiman(2001)}]{breiman_random_2001}
Breiman, L. 2001, 45, 5

\bibitem[{Chollet(2015)}]{chollet_keras_2015}
Chollet, F. 2015.
\newblock \url{https://keras.io}

\bibitem[{Chopin \& Robert(2010)}]{chopin_properties_2010}
Chopin, N., \& Robert, C.~P. 2010, 97, 741, \dodoi{10.1093/biomet/asq021}

\bibitem[{Conselice(2014)}]{conselice_evolution_2014}
Conselice, C.~J. 2014, 52, 291, \dodoi{10.1146/annurev-astro-081913-040037}

\bibitem[{Crawford {et~al.}(1999)Crawford, Allen, Ebeling, Edge, \&
  Fabian}]{crawford_rosat_1999}
Crawford, C.~S., Allen, S.~W., Ebeling, H., Edge, A.~C., \& Fabian, A.~C. 1999,
  306, 857, \dodoi{10.1046/j.1365-8711.1999.02583.x}

\bibitem[{Darg {et~al.}(2010)Darg, Kaviraj, Lintott, Schawinski, Sarzi,
  Bamford, Silk, Andreescu, Murray, Nichol, Raddick, Slosar, Szalay, Thomas, \&
  Vandenberg}]{darg_galaxy_2010}
Darg, D.~W., Kaviraj, S., Lintott, C.~J., {et~al.} 2010, 401, 1552,
  \dodoi{10.1111/j.1365-2966.2009.15786.x}

\bibitem[{Davies {et~al.}(2019)Davies, Serjeant, \&
  Bromley}]{davies_using_2019}
Davies, A., Serjeant, S., \& Bromley, J.~M. 2019, 487, 5263,
  \dodoi{10.1093/mnras/stz1288}

\bibitem[{Donzelli \& Pastoriza(2000)}]{donzelli_spectroscopic_2000}
Donzelli, C.~J., \& Pastoriza, M.~G. 2000, 120, 189, \dodoi{10.1086/301420}

\bibitem[{Drissen {et~al.}(2014)Drissen, Rousseau-Nepton, Lavoie, Robert,
  Martin, Martin, Mandar, \& Grandmont}]{drissen_imaging_2014}
Drissen, L., Rousseau-Nepton, L., Lavoie, S., {et~al.} 2014, 2014, 1,
  \dodoi{10.1155/2014/293856}

\bibitem[{Drissen {et~al.}(2019)Drissen, Martin, Rousseau-Nepton, Robert,
  Martin, Baril, Prunet, Joncas, Thibault, Brousseau, Mandar, Grandmont, Yee,
  \& Simard}]{drissen_sitelle_2019}
Drissen, L., Martin, T., Rousseau-Nepton, L., {et~al.} 2019, 485, 3930,
  \dodoi{10.1093/mnras/stz627}

\bibitem[{Elmegreen {et~al.}(2017)Elmegreen, Elmegreen, Kaufman, Brinks,
  Struck, Bournaud, Sheth, \& Juneau}]{elmegreen_alma_2017}
Elmegreen, D.~M., Elmegreen, B.~G., Kaufman, M., {et~al.} 2017, 841, 43,
  \dodoi{10.3847/1538-4357/aa6ba5}

\bibitem[{Fabbro {et~al.}(2018)Fabbro, Venn, O'Briain, Bialek, Kielty,
  Jahandar, \& Monty}]{fabbro_application_2018}
Fabbro, S., Venn, K., O'Briain, T., {et~al.} 2018, 475, 2978,
  \dodoi{10.1093/mnras/stx3298}

\bibitem[{Foster {et~al.}(2020)Foster, Mendel, Lagos, Wisnioski, Yuan,
  D'Eugenio, Barone, Harborne, Vaughan, Schulze, Remus, Gupta, Collacchioni,
  Khim, Taylor, Bassett, Croom, {McDermid}, Poci, Battisti, Bland-Hawthorn,
  Bellstedt, Colless, Davies, Driver, Ferré-Mateu, Fisher, Gjergo, Johnston,
  Khalid, Kobayashi, Oh, Peng, Robotham, Sweet, Taylor, Tran, Trayford, van~de
  Sande, Yi, \& Zanisi}]{foster_magpi_2020}
Foster, C., Mendel, J.~T., Lagos, C. D.~P., {et~al.} 2020, 2011,
  arXiv:2011.13567.
\newblock \url{http://adsabs.harvard.edu/abs/2020arXiv201113567F}

\bibitem[{González-Gaitán {et~al.}(2019-01-21)González-Gaitán, deSouza,
  Krone-Martins, Cameron, Coelho, Galbany, Ishida, \& {COIN
  collaboration}}]{gonzalez-gaitan_spatial_2019}
González-Gaitán, S., deSouza, R.~S., Krone-Martins, A., {et~al.} 2019-01-21,
  482, 3880, \dodoi{10.1093/mnras/sty2881}

\bibitem[{Hampton {et~al.}(2017)Hampton, Medling, Groves, Kewley, Dopita,
  Davies, Ho, Kaasinen, Leslie, Sharp, Sweet, Thomas, Allen, Bland-Hawthorn,
  Brough, Bryant, Croom, Goodwin, Green, Konstantantopoulos, Lawrence,
  López-Sánchez, Lorente, {McElroy}, Owers, Richards, \&
  Shastri}]{hampton_using_2017}
Hampton, E.~J., Medling, A.~M., Groves, B., {et~al.} 2017, 470, 3395,
  \dodoi{10.1093/mnras/stx1413}

\bibitem[{Hunter(2007)}]{hunter_matplotlib_2007}
Hunter, J.~D. 2007, 9, 90, \dodoi{10.1109/MCSE.2007.55}

\bibitem[{Jenkins \& Peacock(2011)}]{jenkins_power_2011}
Jenkins, C.~R., \& Peacock, J.~A. 2011, 413, 2895,
  \dodoi{10.1111/j.1365-2966.2011.18361.x}

\bibitem[{Kaufman {et~al.}(2012)Kaufman, Grupe, Elmegreen, Elmegreen, Struck,
  \& Brinks}]{kaufman_ngc_2012}
Kaufman, M., Grupe, D., Elmegreen, B.~G., {et~al.} 2012, 144, 156,
  \dodoi{10.1088/0004-6256/144/5/156}

\bibitem[{Keown {et~al.}(2019)Keown, Di~Francesco, Teimoorinia, Rosolowsky, \&
  Chen}]{keown_clover_2019}
Keown, J., Di~Francesco, J., Teimoorinia, H., Rosolowsky, E., \& Chen, M. C.-Y.
  2019, ascl:1909.009.
\newblock \url{http://adsabs.harvard.edu/abs/2019ascl.soft09009K}

\bibitem[{Kewley {et~al.}(2001)Kewley, Heislerr, \&
  Dopita}]{kewley_optical_2001}
Kewley, L., Heislerr, C., \& Dopita, M. 2001, 132, 37

\bibitem[{Kewley {et~al.}(2019)Kewley, Nicholls, \&
  Sutherland}]{kewley_understanding_2019}
Kewley, L.~J., Nicholls, D.~C., \& Sutherland, R.~S. 2019, 57, 511,
  \dodoi{10.1146/annurev-astro-081817-051832}

\bibitem[{Khan {et~al.}(2020)Khan, Sohail, Zahoora, \&
  Qureshi}]{khan_survey_2020}
Khan, A., Sohail, A., Zahoora, U., \& Qureshi, A.~S. 2020, 53, 5455,
  \dodoi{10.1007/s10462-020-09825-6}

\bibitem[{Kieseppä(1997)}]{kieseppa_akaike_1997}
Kieseppä, I.~A. 1997, 48, 21, \dodoi{10.1093/bjps/48.1.21}

\bibitem[{Kim \& Brunner(2017)}]{kim_stargalaxy_2017}
Kim, E.~J., \& Brunner, R.~J. 2017, 464, 4463, \dodoi{10.1093/mnras/stw2672}

\bibitem[{Kreckel {et~al.}(2019)Kreckel, Ho, Blanc, Groves, Santoro,
  Schinnerer, Bigiel, Chevance, Congiu, Emsellem, Faesi, Glover, Grasha,
  Kruijssen, Lang, Leroy, Meidt, {McElroy}, Pety, Rosolowsky, Saito, Sandstrom,
  Sanchez-Blazquez, \& Schruba}]{kreckel_mapping_2019}
Kreckel, K., Ho, I.-T., Blanc, G.~A., {et~al.} 2019, 887, 80,
  \dodoi{10.3847/1538-4357/ab5115}

\bibitem[{Kreckel {et~al.}(2020)Kreckel, Ho, Blanc, Glover, Groves, Rosolowsky,
  Bigiel, Boquíen, Chevance, Dale, Deger, Emsellem, Grasha, Kim, Klessen,
  Kruijssen, Lee, Leroy, Liu, {McElroy}, Meidt, Pessa, Sanchez-Blazquez,
  Sandstrom, Santoro, Scheuermann, Schinnerer, Schruba, Utomo, Watkins, \&
  Williams}]{kreckel_measuring_2020}
---. 2020, 499, 193, \dodoi{10.1093/mnras/staa2743}

\bibitem[{Kuo(2016)}]{kuo_understanding_2016}
Kuo, C.-C.~J. 2016

\bibitem[{Liu {et~al.}(2016)Liu, Shi, Li, Li, Zhu, \& Liu}]{liu_towards_2016}
Liu, M., Shi, J., Li, Z., {et~al.} 2016

\bibitem[{Martin \& Drissen(2017)}]{martin_calibrations_2017}
Martin, T., \& Drissen, L. 2017

\bibitem[{Martin {et~al.}(2012)Martin, Drissen, \& Joncas}]{martin_orbs_2012}
Martin, T., Drissen, L., \& Joncas, G. 2012, in Software and
  Cyberinfrastructure for Astronomy II, Vol. 8451 (International Society for
  Optics and Photonics), 84513K, \dodoi{10.1117/12.925420}

\bibitem[{Martin {et~al.}(2016)Martin, Prunet, \&
  Drissen}]{martin_optimal_2016}
Martin, T.~B., Prunet, S., \& Drissen, L. 2016, 463, 4223,
  \dodoi{10.1093/mnras/stw2315}

\bibitem[{{McKinney}(2010)}]{mckinney_data_2010}
{McKinney}, W. 2010, in Proceedings of the 9th Python in Science Conference,
  Vol.~1 (Scipy Publishing), 56--61, \dodoi{10.25080/Majora-92bf1922-00a}

\bibitem[{Morisset {et~al.}(2015)Morisset, Delgado-Inglada, \&
  Flores-Fajardo}]{morisset_virtual_2015}
Morisset, C., Delgado-Inglada, G., \& Flores-Fajardo, N. 2015, 51, 19

\bibitem[{O'Briain {et~al.}(2021)O'Briain, Ting, Fabbro, Yi, Venn, \&
  Bialek}]{obriain_cycle-starnet_2021}
O'Briain, T., Ting, Y.-S., Fabbro, S., {et~al.} 2021, 906, 130,
  \dodoi{10.3847/1538-4357/abca96}

\bibitem[{Osterbrock \& Ferland(1989)}]{osterbrock_astrophysics_1989}
Osterbrock, D., \& Ferland, G. 1989, Astrophysics of gaseous nebulae and active
  galactic nuclei, 1st edn. (University Science Books)

\bibitem[{Pooley \& Marion(2018)}]{pooley_bayesian_2018}
Pooley, C.~M., \& Marion, G. 2018, 5, 171519, \dodoi{10.1098/rsos.171519}

\bibitem[{Rhea {et~al.}(2020{\natexlab{a}})Rhea, Hlavacek-Larrondo,
  Perreault-Levasseur, Gendron-Marsolais, \& Kraft}]{rhea_novel_2020}
Rhea, C., Hlavacek-Larrondo, J., Perreault-Levasseur, L., Gendron-Marsolais,
  M.-L., \& Kraft, R. 2020{\natexlab{a}}, 160, 202,
  \dodoi{10.3847/1538-3881/abb468}

\bibitem[{Rhea {et~al.}(2021)Rhea, Rousseau-Nepton, Prunet, Prasow-Emond,
  Hlavacek-Larrondo, Vale~Asari, Grasha, \&
  Perreault-Levasseur}]{rhea_machine_2021}
Rhea, C., Rousseau-Nepton, L., Prunet, S., {et~al.} 2021, 2102,
  arXiv:2102.06230.
\newblock \url{http://adsabs.harvard.edu/abs/2021arXiv210206230R}

\bibitem[{Rhea {et~al.}(2020{\natexlab{b}})Rhea, Rousseau-Nepton, Prunet,
  Hlavacek-larrondo, \& Fabbro}]{rhea_machine_2020}
Rhea, C.~L., Rousseau-Nepton, L., Prunet, S., Hlavacek-larrondo, J., \& Fabbro,
  S. 2020{\natexlab{b}}, 901

\bibitem[{Rich {et~al.}(2015)Rich, Kewley, \& Dopita}]{rich_galaxy_2015}
Rich, J.~A., Kewley, L.~J., \& Dopita, M.~A. 2015, 221, 28,
  \dodoi{10.1088/0067-0049/221/2/28}

\bibitem[{Robitaille {et~al.}(2013)Robitaille, Tollerud, Greenfield,
  Droettboom, Bray, Aldcroft, Davis, Ginsburg, Price-Whelan, Kerzendorf,
  Conley, Crighton, Barbary, Muna, Ferguson, Grollier, Parikh, Nair, Günther,
  Deil, Woillez, Conseil, Kramer, Turner, Singer, Fox, Weaver, Zabalza,
  Edwards, Bostroem, Burke, Casey, Crawford, Dencheva, Ely, Jenness, Labrie,
  Lim, Pierfederici, Pontzen, Ptak, Refsdal, Servillat, \&
  Streicher}]{robitaille_astropy_2013}
Robitaille, T.~P., Tollerud, E.~J., Greenfield, P., {et~al.} 2013, 558, A33,
  \dodoi{10.1051/0004-6361/201322068}

\bibitem[{Rousseau-Nepton {et~al.}(2019)Rousseau-Nepton, Martin, Robert,
  Drissen, Amram, Prunet, Martin, Moumen, Adamo, Alarie, Barmby, Boselli,
  Bresolin, Bureau, Chemin, Fernandes, Combes, Crowder, Della~Bruna, Egusa,
  Epinat, Ksoll, Girard, Llanos, Gouliermis, Grasha, Higgs, Hlavacek-Larrondo,
  Ho, Iglesias-Páramo, Joncas, Kam, Karera, Kennicutt, Klessen, Lianou, Liu,
  Liu, de~Amorim, Lyman, Martel, Mazzilli-Ciraulo, {McLeod}, Melchior, Millan,
  Mollá, Momose, Morisset, Pan, Pati, Pellerin, Pellegrini, Pérez, Petric,
  Plana, Rahner, Lara, Sánchez-Menguiano, Spekkens, Stasińska, Takamiya,
  Asari, \& Vílchez}]{rousseau-nepton_signals_2019}
Rousseau-Nepton, L., Martin, R.~P., Robert, C., {et~al.} 2019, 489, 5530,
  \dodoi{10.1093/mnras/stz2455}

\bibitem[{Ruffio {et~al.}(2018)Ruffio, Mawet, Czekala, Macintosh, De~Rosa,
  Ruane, Bottom, Pueyo, Wang, Hirsch, Zhu, \& Nielsen}]{ruffio_bayesian_2018}
Ruffio, J.-B., Mawet, D., Czekala, I., {et~al.} 2018, 156, 196,
  \dodoi{10.3847/1538-3881/aade95}

\bibitem[{Sereno(2016)}]{sereno_bayesian_2016}
Sereno, M. 2016, 455, 2149, \dodoi{10.1093/mnras/stv2374}

\bibitem[{Sharma(2017)}]{sharma_markov_2017}
Sharma, S. 2017, 55, 213, \dodoi{10.1146/annurev-astro-082214-122339}

\bibitem[{Shields(1990)}]{shields_extragalactic_1990}
Shields, G.~A. 1990, 28, 525, \dodoi{10.1146/annurev.aa.28.090190.002521}

\bibitem[{Skilling(2006)}]{skilling_nested_2006}
Skilling, J. 2006, 1, 833, \dodoi{10.1214/06-BA127}

\bibitem[{Soto {et~al.}(2012)Soto, Martin, Prescott, \&
  Armus}]{soto_emission-line_2012}
Soto, K.~T., Martin, C.~L., Prescott, M. K.~M., \& Armus, L. 2012, 757, 86,
  \dodoi{10.1088/0004-637X/757/1/86}

\bibitem[{Speagle(2020)}]{speagle_dynesty_2020}
Speagle, J.~S. 2020, 493, 3132, \dodoi{10.1093/mnras/staa278}

\bibitem[{Struck(2007)}]{struck_star_2007}
Struck, C. 2007, 237, 317, \dodoi{10.1017/S1743921307001664}

\bibitem[{Taaki {et~al.}(2020)Taaki, Kamalabadi, \&
  Kemball}]{taaki_bayesian_2020}
Taaki, J., Kamalabadi, F., \& Kemball, A.~J. 2020, 159, 283,
  \dodoi{10.3847/1538-3881/ab8e38}

\bibitem[{Trotta(2007)}]{trotta_applications_2007}
Trotta, R. 2007, 378, 72, \dodoi{10.1111/j.1365-2966.2007.11738.x}

\bibitem[{Trouille {et~al.}(2013)Trouille, Tremonti, Chen, Crowley-Farenga,
  Rice, \& Loftus}]{trouille_post-starburst_2013}
Trouille, L., Tremonti, C.~A., Chen, Y.-M., {et~al.} 2013, 477, 211.
\newblock \url{http://adsabs.harvard.edu/abs/2013ASPC..477..211T}

\bibitem[{van~der Walt {et~al.}(2011)van~der Walt, Colbert, \&
  Varoquaux}]{van_der_walt_numpy_2011}
van~der Walt, S., Colbert, S.~C., \& Varoquaux, G. 2011, 13, 22,
  \dodoi{10.1109/MCSE.2011.37}

\bibitem[{Van~Rossum \& Drake(2009)}]{van_rossum_python_2009}
Van~Rossum, G., \& Drake, F.~L. 2009, Python 3 Reference Manual (Create Soace)

\bibitem[{Veilleux \& Osterbrock(1987)}]{veilleux_spectral_1987}
Veilleux, S., \& Osterbrock, D.~E. 1987, 63, 295

\bibitem[{Virtanen {et~al.}(2020)Virtanen, Gommers, Oliphant, Haberland, Reddy,
  Cournapeau, Burovski, Peterson, Weckesser, Bright, van~der Walt, Brett,
  Wilson, Millman, Mayorov, Nelson, Jones, Kern, Larson, Carey, Polat, Feng,
  Moore, {VanderPlas}, Laxalde, Perktold, Cimrman, Henriksen, Quintero, Harris,
  Archibald, Ribeiro, Pedregosa, van Mulbregt, \&
  Contributors}]{virtanen_scipy_2020}
Virtanen, P., Gommers, R., Oliphant, T.~E., {et~al.} 2020, 17, 261,
  \dodoi{10.1038/s41592-019-0686-2}

\bibitem[{Waskom {et~al.}(2017)Waskom, Botvinnik, O'Kane, Hobson, Lukauskas,
  Gemperline, Augspurger, Halchenko, Cole, Warmenhoven, de~Ruiter, Pye, Hoyer,
  Vanderplas, Villalba, Kunter, Quintero, Bachant, Martin, Meyer, Miles, Ram,
  Yarkoni, Williams, Evans, Fitzgerald, Brian, Fonnesbeck, Lee, \&
  Qalieh}]{michael_waskom_mwaskomseaborn_2017}
Waskom, M., Botvinnik, O., O'Kane, D., {et~al.} 2017,
  \dodoi{10.5281/zenodo.883859}

\bibitem[{Wild {et~al.}(2009)Wild, Walcher, Johansson, Tresse, Charlot, Pollo,
  Le~Fèvre, \& De~Ravel}]{wild_post-starburst_2009}
Wild, V., Walcher, C.~J., Johansson, P.~H., {et~al.} 2009, 395, 144,
  \dodoi{10.1111/j.1365-2966.2009.14537.x}

\end{thebibliography}
\bibliographystyle{aasjournal}

\end{document}